\documentclass[manuscript,nonacm]{acmart}

\AtBeginDocument{%
  }

\begin{document}

\title{%From Online Advice to AI: 
Postpartum Sense-Making and AI in a Fragmented Care System}
\author{Diana Freed}
\authornote{Both authors contributed equally to this work.}
\email{diana\_freed@brown.edu}
\affiliation{%
  \institution{Brown University}
  \city{Providence}
  \state{Rhode Island}
  \country{USA}
}

\author{Maja Mishevska}
\authornotemark[1]
\email{maja\_mishevska@brown.edu}
\affiliation{%
  \institution{Brown University}
  \city{Providence}
  \state{Rhode Island}
  \country{USA}
}

\begin{abstract}
Postpartum women must navigate health information across a fragmented landscape of clinicians, peers, online communities, influencers, and, increasingly, large language models (LLMs). Existing evaluations of LLMs in health contexts primarily ask whether systems provide accurate, safe, and readable information. We argue that these criteria, while essential, are insufficient for understanding how LLMs may affect postpartum care. LLMs enter an information ecology in which women are already required to reconcile competing forms of expertise, determine whether their experiences are normal or concerning, and decide when and where to seek care. At the same time, online representations of motherhood and femininity shape which postpartum experiences are made visible and what forms of self-management are expected of women. Drawing together research on postpartum care gaps, online help-seeking, influencer cultures, gendered responsibility, and AI-supported health information, we position LLMs as emerging intermediaries within this broader care infrastructure. We argue that evaluating postpartum AI requires examining not only information quality, but also how these systems shape epistemic coordination, transitions to formal care, and the distribution of responsibility for making fragmented systems of care work.

\end{abstract}


\maketitle

\noindent\textbf{Workshop:}
CSCW 2026 -- Women's Health as Cooperative Work:
Participatory AI Across Care, Community, and Policy
(October 11, 2026).

\par\smallskip
\noindent\textbf{Proceedings index:}
\url{https://arxiv.org/abs/2611.03776}

\par\smallskip
\noindent\textbf{Assigned report number:}
\texttt{CSCW-WHCW-2026-06}

\par\medskip

\section{Introduction}

Globally, it is estimated that at least 40 million women each year experience long-term health problems related to childbirth, spanning physical and mental health conditions \cite{vogel2024postpartum}. 
Postpartum women\footnote{We use \emph{postpartum women} because this paper focuses on gendered expectations around motherhood and femininity, while recognizing that postpartum experiences occur across gender identities, including among trans, nonbinary, and other gender-diverse people.} must continually interpret their experiences, decide when to seek help, and determine which sources of guidance to trust.

These decisions are made within a care system marked by gaps in preparation, continuity, and support \cite{tully2017fourth,nazarenko2024preparation,halm2024clinicians}. As a result, women often look outside clinical care to a broader information environment that includes social-support networks, peers, online communities, and influencers \cite{baker2018socialsupport,chivers2021supportseeking}. General-purpose large language models (LLMs) such as ChatGPT, a form of generative artificial intelligence (AI), are emerging within this same landscape, offering a conversational way to seek health information and interpret personal experiences \cite{alam2024chatgpt}.

Recent public cases illustrate the stakes of postpartum interpretation. Earlier in 2026, actress Hayden Panettiere reflected on the consequences she faced after publicly disclosing postpartum depression (PPD). She said that longtime sponsor Neutrogena sought to end its relationship with her after she spoke publicly about the condition, an experience that made clear to her ``how misunderstood it is, how much stigma there is around it'' \cite{obrien2026postpartum}. Following her death, that account resurfaced across media.

Additionally, in the highly publicized Lindsay Clancy legal case, the defense team for a Massachusetts mother argued that she was experiencing postpartum psychosis at the time of her three children's deaths \cite{obrien2026postpartum,raymond2026clancy}. \footnote{The trial ended in a mistrial in September 2026 after the jury was unable to reach a unanimous verdict \cite{axios2026clancy}.} Although these cases are profoundly different, both drew public attention to a persistent challenge: postpartum experiences may be misunderstood, stigmatized, dismissed, or recognized only after substantial harm has occurred. These challenges are compounded by fragmented systems of care, in which women may struggle not only to recognize when they need help, but also to determine where to seek it and whether their concerns will be taken seriously.
 
We argue that evaluating general-purpose LLMs in postpartum health requires moving beyond the quality of individual answers to examine how these systems participate in postpartum sense-making. When used for postpartum health questions, LLMs enter an already fragmented care environment and may shape how experiences are interpreted, when concerns are treated as warranting care, and how women navigate between informal and formal sources of support. Their value therefore depends not only on factual accuracy and safety, but on whether they reduce or reproduce the interpretive and coordination work already placed on women.

\section{Postpartum Care Gaps and Interpretive Work}

\paragraph{Fragmented care and coordination.} The early postpartum period---often described as the ``fourth trimester''---involves substantial physical, psychological, and social adjustment \cite{tully2017fourth}. Postpartum needs span physical recovery, mental health, feeding, sleep, relationships, and caregiving, yet support is often spread across different providers and settings. Women report feeling underprepared for the postpartum period, while clinicians have identified limitations in the continuity and structure of existing care \cite{tully2017fourth,nazarenko2024preparation,halm2024clinicians}. In practice, this can leave women acting as de facto coordinators of their own postpartum care, connecting information across providers and deciding where, when, and from whom to seek support.

\paragraph{Help-seeking and PPD} Help-seeking therefore depends on more than the availability of services: concerns must first be recognized as warranting care, and women must feel able to disclose them. Research on perinatal mental-health help-seeking identifies barriers at multiple levels, including stigma, fear of judgment, limited knowledge and trust, practical constraints such as cost and childcare, and structural limitations in the availability or responsiveness of care \cite{hu2023helpseeking,daehn2022literacy}. These dynamics are not experienced equally: racial and ethnic inequities persist in postpartum diagnosis and care, while Black birthing people report discriminatory treatment, distrust, and limited cultural responsiveness \cite{haight2024inequities,segovia2025blackbirthing}.

PPD is a particularly important subset of postpartum health: a global meta-analysis of 565 studies across 80 countries estimated its prevalence at approximately 17\% \cite{wang2021ppdprevalence}. Its temporal boundaries also vary across clinical frameworks: the DSM-5 ``with peripartum onset'' specifier uses a four-week postpartum cutoff, which has been criticized as excluding later-onset episodes \cite{sharma2014peripartum}, whereas ACOG recognizes perinatal mental-health conditions through the first year postpartum \cite{acog2023mentalhealth}. This definitional variability may further complicate help-seeking for women experiencing PPD.

\paragraph{Normalization and pathologization.} These barriers and uneven experiences of care complicate the boundary between normalization and pathologization. Reassurance that an experience is ``normal postpartum'' may be valuable, but it can also minimize concerns that warrant attention, while ordinary variation may be medicalized as dysfunction requiring correction. Online postpartum discussions reflect this tension: women negotiate medicalized interpretations of their experiences, and peer normalization can sometimes discourage appropriate care \cite{joshua2026postpartumproof,rouhi2021normalcy}.

Similar tensions appear across women’s reproductive health, where experiences may be medicalized through healthcare institutions, industry, and media, while symptoms are also dismissed or subject to ``medical gaslighting'' \cite{kirli2025medicalization,kokab2026gaslighting}. Because such misrecognition can delay diagnosis and worsen outcomes, the challenge is determining how competing accounts of what is normal or pathological should be interpreted and acted upon.

\section{Postpartum Sense-Making in Online Information Ecologies}
\subsection{Informal and Online Sources of Support}

Online spaces are deeply embedded in how new mothers seek information and support. In one study, 89\% of mothers reported using social media for pregnancy or parenting advice, and 84\% considered social-media contacts a source of support \cite{baker2018socialsupport}. More recently, a 2026 national survey found that 51\% of women ages 18--44 would first turn to an AI chatbot, social media, or a search engine for an urgent breastfeeding concern rather than a doctor or family member \cite{orlandohealth2026breastfeeding}. Beyond obtaining information, postpartum forums offer reassurance and peer exchange, helping women compare experiences, test possible solutions, and build confidence \cite{chivers2021supportseeking}. Yet access to more information does not necessarily make postpartum experiences easier to interpret; women also report anxiety and information overload when navigating online health information \cite{barkin2016information}.

Online platforms also make some aspects of postpartum more visible than others. An analysis of 700 \#postpartum Instagram posts found exercise to be the most common content category, alongside emotions, parenting, relationships, and the maternal body \cite{macpherson2023postpartuminstagram}. Similar tensions appear elsewhere in women’s health: women searching for menopause information online often report unmet questions and consult multiple sources to verify what they find, while social-media content in the same space is heavily shaped by commercial promotion, including branded supplements \cite{funnell2025menopausesearch,eubanks2026menopausesupplements}. Women are therefore navigating an information environment in which clinical, experiential, and commercial accounts are difficult to separate, shifting additional work of verification and interpretation onto them.

\subsection{Momfluencers and Postpartum Norms}

Within this environment, health information can also carry expectations about how mothers should behave. Breastfeeding is one example: it is closely tied to ideas of ``good motherhood,'' while perceived pressure to breastfeed has been associated with poorer postpartum mental-health outcomes \cite{marshall2007goodmother,grattan2024breastfeedingpressure}. Mothers engaging with breastfeeding content also weigh influencer advice against their own experiences and other sources of knowledge \cite{mertens2026breastfeeding}. Similar pressures appear around bodily recovery. ``Bounce back'' culture frames postpartum recovery around returning to a pre-pregnancy body, and exposure to \#fitspo (fitness-inspiration) content has been linked to greater upward social comparison and anxiety and lower body satisfaction, while body-positive content can have the opposite effect \cite{wenhold2025bopo}.

 Momfluencers---social media creators who build audiences around motherhood and parenting---occupy a particularly complex position because they combine personal experience, advice, role modeling, and commercial influence. While they can provide support and shared experience, larger accounts tend to present more marketable and aspirational versions of motherhood, with maternal health issues relatively rarely foregrounded \cite{luckbenz2026motherhood}. As creators professionalize, platform incentives around visibility, engagement, and monetization may further favor socially acceptable portrayals over more complex experiences. Momfluencers can therefore shape both what women learn about postpartum life and which versions of motherhood become most visible. At the same time, some may themselves be navigating the postpartum period, and the pressure to sustain a curated, marketable persona may constrain what they feel able to share publicly about their own health and difficulties.

% Their content can include breastfeeding guidance, postpartum recovery narratives, everyday parenting routines, and product promotion \cite{luckbenz2026motherhood,mertens2026breastfeeding,gibson2025mumfluencers}. 

\subsection{Traditional Femininity and Gendered Responsibility}

Platform logics can broaden exposure beyond mainstream content: engagement with parenting and domestic-life material may lead recommendation systems to surface adjacent narratives of femininity that users did not deliberately seek out \cite{sykes2024tradwives}. This can include ``tradwife'' content, which curates and monetizes idealized versions of wifehood and homemaking rooted in traditional gender roles. Presented through everyday lifestyle content, these representations can normalize restrictive gender hierarchies and carry anti-feminist and broader right-wing ideas about gender and family \cite{sykes2024tradwives,miller2026submissivewife}. Scholars have situated their appeal partly in material pressures that are also salient after childbirth, including weak care systems and the devaluation of reproductive labor, with domestic retreat and individualized family care offered as an apparent alternative \cite{banetweiser2026tradwives}.

What remains largely unexplored is how these gendered lifestyle narratives intersect with postpartum health information in an evolving digital ecology. This matters because tradwife content can include anti-scientific and anti-institutional health discourse framed through ``natural'' alternatives and distrust of mainstream expertise \cite{scott2025tradwives}. Through the lens of \emph{gendered consumer responsibilization}, women may also be positioned as self-managing subjects responsible for interpreting and acting on health advice in ways shaped by expectations of femininity \cite{corus2025responsibilization}. We therefore treat tradwife content as a potentially important part of the normative environment shaping postpartum sense-making, without implying that exposure leads directly to ideological adoption.

% This is particularly consequential when advice about feeding, recovery, caregiving, or ``natural'' living is interwoven with broader expectations about how mothers should behave and care for themselves and others.
General-purpose LLMs introduce a different dynamic into this information ecology: rather than presenting these forms of knowledge separately, they can synthesize clinical, experiential, commercial, and normative accounts into a single, coherent-seeming response.

\section{LLMs as Emerging Postpartum Intermediaries}

Prior research on AI applications in postpartum health has often focused on machine learning and related predictive methods for detecting or estimating risk of conditions such as PPD \cite{alkhateeb2026ppdai,xia2025ppdprediction}. Separate work has explored clinically situated conversational agents that support postpartum patients after discharge or respond to maternal-health questions while facilitating screening and follow-up when concerns arise \cite{leitner2025conversational,peters2026chatbot}. Unlike these more structured tools, general-purpose LLMs such as ChatGPT, Claude, and Gemini are available outside clinical settings and can support open-ended health-related conversation. This format may allow users to describe experiences in their own words and receive immediate, personalized-seeming guidance, which may feel particularly accessible for concerns that are uncertain or difficult to disclose \cite{alam2024chatgpt}.

Research evaluating general-purpose LLMs in postpartum health remains limited. Among the studies considered here, a recent evaluation of AI chatbots answering PPD questions found high accuracy for some LLMs, with ChatGPT-5 achieving 97.5\% accuracy on standardized multiple-choice questions \cite{yang2026ppdchatbots}. The same study, however, identified limitations in transparency and source attribution and did not assess clinical safety, patient comprehension, user behavior, or real-world effectiveness. Other women's-health studies similarly report variation in LLM accuracy and readability \cite{karam2026menopausellm}, while broader frameworks consider dimensions including safety, consensus, objectivity, reproducibility, and explainability \cite{deva2025menopausellm}. Together, this work suggests that accuracy is an important dimension of evaluation while leaving open how LLMs may shape help-seeking in practice.

\section{From Information Quality to Care Infrastructure}

Treating LLMs as standalone information tools risks reproducing the individualization already present in postpartum care: women may receive more information while remaining responsible for interpreting, verifying, and acting on it. Recent sociotechnical work on AI-supported peripartum information-seeking similarly argues for inspectable governance, pluralistic verification, recourse, and integration with existing care without shifting additional burden onto patients \cite{balloli2026sociotechnical}. We therefore examine three dimensions of LLM-supported postpartum care: epistemic coordination, help-seeking and care transitions, and responsibility and accountability.

\paragraph{Epistemic coordination.} LLMs can synthesize clinical guidance, peer experience, commercial content, and gendered narratives into a single response. In doing so, they may obscure differences in origin, evidence, and authority, while also participating in the normalization or medicalization of postpartum experiences. 

Evaluation should therefore examine how systems communicate uncertainty and how they frame experiences as normal, concerning, or pathological. For CSCW, this also raises a relational question: whose interpretation becomes authoritative, and what happens when a woman’s account does not align with it? A system could help articulate a concern that has been difficult to communicate, or provide another interpretation that one must defend, contest, or reconcile. We propose studying these possibilities as part of the social work of establishing what a concern means.
%rather than reducing epistemic coordination to the presence of source citations.

\paragraph{Help-seeking and care transitions.} 
LLMs may become another step between informal sense-making and formal care, offering reassurance, self-management guidance, or recommendations to seek professional support. The question is whether they help women navigate transitions across sources of care and reduce coordination work or simply add another layer that women must reconcile themselves. Studying these trajectories can reveal where information is preserved, transformed, or lost and identify opportunities for AI to support continuity across people and settings.

\paragraph{Responsibility and accountability.} 
%LLMs may reinforce gendered responsibilization by making individual self-management more efficient without necessarily reducing the work placed on women \cite{corus2025responsibilization}. 
%Postpartum AI should therefore be evaluated by whether it redistributes---or reproduces---the labor and responsibility required to make care function.
LLMs may reinforce gendered responsibilization by making individual self-management more efficient without necessarily reducing the work placed on women \cite{corus2025responsibilization}. This concern is compounded when systems obscure the limits of their guidance, the basis for their recommendations, or where responsibility lies when advice is incomplete or cannot be acted upon. Transparency should therefore extend beyond identifying information sources to making clear what an LLM can and cannot determine, when additional expertise is needed, and who remains responsible for follow-up. 

%An LLM may help an individual interpret symptoms, compare recommendations, prepare questions for a clinician, or identify next steps while still leaving her responsible for ensuring that care actually happens. 
Making care work easier to perform is not the same as redistributing that work. Postpartum AI should therefore be evaluated by whether it makes these boundaries of responsibility visible and reduces, redistributes, or reproduces the labor required to make care function.

\section{Conclusion}
LLMs offer meaningful possibilities for making postpartum information more immediate and accessible, and for improving its factual accuracy and safety. But accurate answers alone cannot repair a poorly coordinated care environment. Postpartum AI should also be evaluated by how it shapes interpretation, trust, help-seeking, responsibility, and movement across sources of care. There is an opportunity for AI-supported systems to contribute to more coordinated, comprehensive, and accountable forms of postpartum care.

%% Bibliography
\bibliographystyle{ACM-Reference-Format}
\bibliography{references}

@misc{obrien2026postpartum,
  author       = {Sara Ashley O'Brien},
  title        = {A Celebrity Death and a Murder Trial Shine Light on Postpartum Mental Health},
  year         = {2026},
  month        = aug,
  howpublished = {The Wall Street Journal},
  note         = {Published August 21, 2026},
  url          = {https://www.wsj.com/articles/hayden-panettiere-lindsay-clancy-postpartum-mental-health-397b6da1}
}

@misc{raymond2026clancy,
  author       = {Nate Raymond and Maria Alejandra Cardona and Luc Cohen},
  title        = {Lindsay Clancy trial over killing her children ends in mistrial},
  year         = {2026},
  month        = sep,
  howpublished = {Reuters},
  note         = {Published September 4, 2026},
  url          = {https://www.reuters.com/world/us/deadlocked-jury-lindsay-clancys-murder-trial-deliberate-seventh-day-2026-09-04/}
}

@article{vogel2024postpartum,
  author  = {Joshua P. Vogel and Jenny Jung and Tina Lavin and Grace Simpson
             and Dvora Kluwgant and Edgardo Abalos and Virginia Diaz
             and Soo Downe and Veronique Filippi and Ioannis Gallos
             and Hadiza Galadanci and Geetanjali Katageri
             and Caroline S. E. Homer and G. Justus Hofmeyr
             and Tippawan Liabsuetrakul and Imran O. Morhason-Bello
             and Alfred Osoti and Jo{\~a}o Paulo Souza and Ranee Thakar
             and Shakila Thangaratinam and Olufemi T. Oladapo},
  title   = {Neglected medium-term and long-term consequences of labour and childbirth:
             a systematic analysis of the burden, recommended practices,
             and a way forward},
  journal = {The Lancet Global Health},
  year    = {2024},
  volume  = {12},
  number  = {2},
  pages   = {e317--e330},
  doi     = {10.1016/S2214-109X(23)00454-0}
}

@article{kirli2025medicalization,
  author  = {Gamze K{\i}rl{\i} and {\c{S}}erife Didem Kaya},
  title   = {Medicalization of female life stages: a qualitative research},
  journal = {BMC Health Services Research},
  year    = {2025}, volume = {25}, number = {1}, pages = {322},
  doi     = {10.1186/s12913-024-12181-8}
}

@article{tully2017fourth,
  author  = {Kristin P. Tully and Alison M. Stuebe and Sarah B. Verbiest},
  title   = {The fourth trimester: a critical transition period with unmet maternal health needs},
  journal = {American Journal of Obstetrics and Gynecology},
  year    = {2017}, volume = {217}, number = {1}, pages = {37--41},
  doi     = {10.1016/j.ajog.2017.03.032}
}

@article{kokab2026gaslighting,
  author  = {Farina Kokab and Stephen J. Williams and Niha Hussain and Nicola Gale},
  title   = {From `dismissal' to `medical gaslighting': A literature review of women's experiences of reproductive health services in the {UK}},
  journal = {Midwifery},
  year    = {2026}, volume = {159}, pages = {104823},
  doi     = {10.1016/j.midw.2026.104823}
}

@article{nazarenko2024preparation,
  author  = {Danielle N. Nazarenko and Ariel L. Daniel and Stephanie Durfee and Kafuli Agbemenu},
  title = {Parent-identified gaps in preparation for the postpartum period in the {United States}: An integrative review},
  journal = {Birth},
  year    = {2024}, volume = {51}, number = {4}, pages = {669--689},
  doi     = {10.1111/birt.12832}
}

@article{halm2024clinicians,
  author  = {Abby E. Halm and Paige Hornsby and Shreya Shetty and Cristalle Madray and Ann Kellams},
  title   = {Clinicians Speak: We Need to Redesign Postpartum Care},
  journal = {The Journal of Perinatal Education},
  year    = {2024}, volume = {33}, number = {3}, pages = {159--171},
  doi     = {10.1891/JPE-2023-0023}
}

@article{hu2023helpseeking,
  author  = {Ying Hu and Sasa Huang and Meili Xiao and Bing Fu and Guanxiu Tang and Lisa Lommel and Jun Lei},
  title   = {Barriers and facilitators of psychological help-seeking behaviors for perinatal women with depressive symptoms: A qualitative systematic review based on the Consolidated Framework for Implementation Research},
  journal = {Midwifery},
  year    = {2023}, volume = {122}, pages = {103686},
  doi     = {10.1016/j.midw.2023.103686}
}

@article{daehn2022literacy,
  author  = {Daria Daehn and Sophie Rudolf and Silke Pawils and Babette Renneberg},
  title   = {Perinatal mental health literacy: knowledge, attitudes, and help-seeking among perinatal women and the public -- a systematic review},
  journal = {BMC Pregnancy and Childbirth},
  year    = {2022}, volume = {22}, number = {1}, pages = {574},
  doi     = {10.1186/s12884-022-04865-y}
}

@article{haight2024inequities,
  author  = {Sarah C. Haight and Jamie R. Daw and Chantel L. Martin and Karen Sheffield-Abdullah and Sarah Verbiest and Brian W. Pence and Joanna Maselko},
  title = {Racial and Ethnic Inequities in Postpartum Depressive Symptoms, Diagnosis, and Care in 7 {US} Jurisdictions},
  journal = {Health Affairs},
  year    = {2024}, volume = {43}, number = {4}, pages = {486--495},
  doi     = {10.1377/hlthaff.2023.01434}
}

@article{segovia2025blackbirthing,
  author  = {Laura M. Segovia and Emily Neiman and Shannon L. Gillespie and McKenzie K. Jancsura and Cindy M. Anderson},
  title   = {Prenatal and Postpartum Care Experiences Among Black Birthing People in the United States: An Integrative Review},
  journal = {Journal of Midwifery \& Women's Health},
  year    = {2025}, volume = {70}, number = {2}, pages = {235--246},
  doi     = {10.1111/jmwh.13705}
}

@article{marshall2007goodmother,
  author  = {Joyce L. Marshall and Mary Godfrey and Mary J. Renfrew},
  title   = {Being a `good mother': managing breastfeeding and merging identities},
  journal = {Social Science \& Medicine},
  year    = {2007}, volume = {65}, number = {10}, pages = {2147--2159},
  doi     = {10.1016/j.socscimed.2007.06.015}
}

@article{grattan2024breastfeedingpressure,
  author  = {Rebecca E. Grattan and Sophie M. London and Georgia E. Bueno},
  title   = {Perceived pressure to breastfeed negatively impacts postpartum mental health outcomes over time},
  journal = {Frontiers in Public Health},
  year    = {2024}, volume = {12}, pages = {1357965},
  doi     = {10.3389/fpubh.2024.1357965}
}

@article{baker2018socialsupport,
  author  = {Brenda Baker and Irene Yang},
  title   = {Social media as social support in pregnancy and the postpartum},
  journal = {Sexual \& Reproductive Healthcare},
  year    = {2018}, volume = {17}, pages = {31--34},
  doi     = {10.1016/j.srhc.2018.05.003}
}

@article{macpherson2023postpartuminstagram,
  author  = {Ashley R. MacPherson and Morgan Reid and Natalie Dautovich},
  title   = {Examining the postpartum period through social media: A content and thematic analysis of {\#}postpartum {Instagram} posts},
  journal = {Psychology of Popular Media},
  year    = {2023}, volume = {12}, number = {4}, pages = {424--433},
  doi     = {10.1037/ppm0000422}
}

@article{eubanks2026menopausesupplements,
  author  = {Allison A. Eubanks and Katerina Shvartsman},
  title   = {Social Media and Supplements for Menopausal Symptoms: A Content Analysis},
  journal = {BJOG: An International Journal of Obstetrics \& Gynaecology},
  year    = {2026}, volume = {133}, number = {10}, pages = {1885--1892},
  doi     = {10.1111/1471-0528.70242}
}

@article{funnell2025menopausesearch,
  author  = {Erin Lucy Funnell and Freya McConnell and Nayra A. Martin-Key and Leyao Qian and Kathryn Babbitt and Sabine Bahn},
  title   = {Understanding Experiences of and Unmet Needs in Online Searches for Menopause Information: An Exploratory Survey},
  journal = {JMIR Formative Research},
  year    = {2025}, volume = {9}, pages = {e75335},
  doi     = {10.2196/75335}
}

@article{chivers2021supportseeking,
  author  = {Bonnie R. Chivers and Rhonda M. Garad and Lisa J. Moran and Siew Lim and Cheryce L. Harrison},
  title   = {Support Seeking in the Postpartum Period: Content Analysis of Posts in Web-Based Parenting Discussion Groups},
  journal = {Journal of Medical Internet Research},
  year    = {2021},
  volume  = {23},
  number  = {7},
  pages   = {e26600},
  doi     = {10.2196/26600}
}

@article{joshua2026postpartumproof,
  author  = {Kaitlin H. Joshua},
  title   = {``I thought I could postpartum-proof my experience'': Embodying, resisting, and negotiating medicalization online},
  journal = {Social Science \& Medicine},
  year    = {2026}, volume = {398}, pages = {119157},
  doi     = {10.1016/j.socscimed.2026.119157}
}

@article{rouhi2021normalcy,
  author  = {Maryam Rouhi and Christine Stirling and Elaine Peta Crisp},
  title   = {The `fallacy of normalcy': A content analysis of women's online post-childbirth health-related support},
  journal = {Women and Birth},
  year    = {2021}, volume = {34}, number = {3}, pages = {e262--e270},
  doi     = {10.1016/j.wombi.2020.04.007}
}

@article{barkin2016information,
  author  = {Jennifer L. Barkin and Smit Jani},
  title   = {Information Management in New Motherhood: Does the Internet Help or Hinder?},
  journal = {Journal of the American Psychiatric Nurses Association},
  year    = {2016}, volume = {22}, number = {6}, pages = {475--482},
  doi     = {10.1177/1078390316659697}
}

@article{luckbenz2026motherhood,
  author  = {Julia L{\"u}ck-Benz and Claudia Wegener and Shari Adlung and Friederike Jage-D'Aprile},
  title   = {Motherhood curated: A content analysis of momfluencer representations by audience size},
  journal = {European Journal of Communication},
  year    = {2026}, volume = {41}, number = {3}, pages = {226--246},
  doi     = {10.1177/02673231261416007}
}

@article{mertens2026breastfeeding,
  author  = {Ellen Mertens and Emma Beuckels and Liselot Hudders},
  title   = {Negotiating breastfeeding information in everyday life: mothers' engagement with momfluencer content},
  journal = {International Journal of Advertising},
  year    = {2026}, volume = {45}, number = {6}, pages = {1475--1498},
  doi     = {10.1080/02650487.2026.2707851}
}

@article{wenhold2025bopo,
  author  = {Halie Wenhold and Amelia C. Couture Bue and Ciera E. Kirkpatrick},
  title   = {{\#}Bopo or Bounce Back?: Investigating the Impact of Social Media Videos on Postpartum Mothers},
  journal = {Health Communication},
  year    = {2025}, volume = {40}, number = {12}, pages = {2674--2685},
  doi     = {10.1080/10410236.2025.2475856}
}

@article{corus2025responsibilization,
  author  = {Canan Corus and Bige Saatcioglu and Ozlem Sandikci},
  title   = {Gendered Consumer Responsibilization: The Constitution of Menopausal Women as Responsible Feminine Consumer Subjects},
  journal = {Journal of Consumer Research},
  year    = {2025}, volume = {52}, number = {2}, pages = {351--371},
  doi     = {10.1093/jcr/ucae057}
}

@article{sykes2024tradwives,
  author  = {Sophia Sykes and Veronica Hopner},
  title   = {Tradwives: Right-Wing Social Media Influencers},
  journal = {Journal of Contemporary Ethnography},
  year    = {2024}, volume = {53}, number = {4}, pages = {453--487},
  doi     = {10.1177/08912416241246273}
}

@article{miller2026submissivewife,
  author  = {Elizabeth A. Miller and Elizabeth R. Cole},
 title = {Selling the {\#SubmissiveWife}: How White Tradwife Influencers Commodify Traditional Femininity},
  journal = {Psychology of Women Quarterly},
  year    = {2026}, volume = {50}, number = {2}, pages = {149--164},
  doi     = {10.1177/03616843261434820}
}

@article{scott2025tradwives,
  author  = {Kate Scott and Lindsay Day},
  title   = {{TikTok} tradwives: femininity, reproduction, and social media},
  journal = {Gender and Education},
  year    = {2026},
  volume  = {38},
  number  = {6},
  pages   = {716--733},
  doi     = {10.1080/09540253.2025.2546050}
}

@article{banetweiser2026tradwives,
  author  = {Sarah Banet-Weiser and Sara Reinis},
  title   = {The rage of tradwives: Affective economies and romanticizing retreat},
  journal = {Feminist Theory},
  year    = {2026},
  volume  = {27},
  number  = {1},
  pages   = {50--66},
  doi     = {10.1177/14647001251371971}
}

@article{alam2024chatgpt,
  author  = {Sultan Alam and Shahab Saquib Sohail},
  title   = {Integrating {ChatGPT}: Enhancing postpartum mental healthcare with artificial intelligence ({AI}) support},
  journal = {Digital Health},
  year    = {2024}, volume = {10}, pages = {20552076241295565},
  doi     = {10.1177/20552076241295565}
}

@article{leitner2025conversational,
  author  = {Kirstin Leitner and Clare Cutri-French and Abigail Mandel and Lori Christ and N. Koelper and M. McCabe and E. Seltzer and L. Scalise and J. A. Colbert and Anuja Dokras and R. Rosin and Lisa Levine},
  title   = {A Conversational Agent Using Natural Language Processing for Postpartum Care for New Mothers: Development and Engagement Analysis},
  journal = {JMIR AI},
  year    = {2025}, volume = {4}, pages = {e58454},
  doi     = {10.2196/58454}
}

@article{peters2026chatbot,
  author  = {Carson J. Peters and Valerie Aldana Lainez and Kaili Clark and Michelle Jasczynski and Quynh C. Nguyen and Elizabeth M. Norell},
  title   = {Using an {AI}-powered Mobile Application Chatbot to Address Maternal Depression Indicators and Inquiries in the Perinatal and Postpartum Periods: A Multimethod Analysis},
  journal = {INQUIRY: The Journal of Health Care Organization, Provision, and Financing},
  year    = {2026}, volume = {63}, pages = {00469580261417580},
  doi     = {10.1177/00469580261417580}
}

@misc{orlandohealth2026breastfeeding,
  author       = {{Orlando Health}},
  title        = {The Real World Risk of A.I. Breastfeeding Advice},
  year         = {2026},
  month        = aug,
  howpublished = {Orlando Health Multimedia Newsroom},
  note         = {Published August 17, 2026},
  url          = {https://oh.multimedianewsroom.org/2026/08/17/the-real-world-risk-of-a-i-breastfeeding-advice-2/}
}

@article{yang2026ppdchatbots,
  author  = {Man Yang and Hui Liu and Shuyan Lin and Linyan Liu and Yunxia Wang and Xiuling Qiu and Ling Wei and Li Zhou and Xinxin Wang},
  title   = {Evaluating the accuracy, reliability, and readability of {AI} chatbots in delivering postpartum depression information},
  journal = {Frontiers in Psychiatry},
  year    = {2026}, volume = {17}, pages = {1881536},
  doi     = {10.3389/fpsyt.2026.1881536}
}

@article{karam2026menopausellm,
  author  = {Jana Karam and Chrisandra Shufelt and Nancy Safwan and Ekta Kapoor and Monica M. Christmas and Stephanie S. Faubion},
  title   = {Evaluation of the accuracy and readability of large language model responses on menopause and hormone therapy},
  journal = {Menopause},
  year    = {2026}, volume = {33}, number = {5}, pages = {515--520},
  doi     = {10.1097/GME.0000000000002695}
}

@inproceedings{deva2025menopausellm,
  author    = {Roshini Deva and Manvi S and Jasmine Zhou and Elizabeth Britton Chahine and Agena Davenport-Nicholson and Nadi Nina Kaonga and Selen Bozkurt and Azra Ismail},
  title     = {A Mixed-Methods Evaluation of {LLM}-Based Chatbots for Menopause},
  booktitle = {Intelligent Health Systems -- From Technology to Data and Knowledge: Proceedings of MIE 2025},
  series    = {Studies in Health Technology and Informatics},
  year      = {2025}, volume = {327}, pages = {1175--1179},
  doi       = {10.3233/SHTI250575}
}

@article{alkhateeb2026ppdai,
  author  = {Mais Alkhateeb and Ajisha Nayeem and Arfan Ahmed and Mohammed Alsahli and Javaid Sheikh and Alaa Abd-Alrazaq},
  title   = {{AI} for Detecting and Predicting Postpartum Depression: Scoping Review},
  journal = {Journal of Medical Internet Research},
  year    = {2026}, volume = {28}, pages = {e77376},
  doi     = {10.2196/77376}
}

@inproceedings{balloli2026sociotechnical,
  author    = {Vaibhav Balloli and Julia Erickson and Xinyi Li and Erin MacMurray van Liemt and Alex Peahl and Elizabeth Bondi-Kelly},
  title     = {``Where is this coming from?'' Uncovering Trustworthiness Ideals in {AI}-powered Peripartum Information Seeking},
  booktitle = {Proceedings of the 2026 ACM Conference on Fairness, Accountability, and Transparency},
  year      = {2026},
  pages     = {1979--2009},
  publisher = {Association for Computing Machinery},
  doi       = {10.1145/3805689.3812277},
  url       = {https://doi.org/10.1145/3805689.3812277}
}

@article{xia2025ppdprediction,
  author  = {Jie Xia and Chen Chen and Xiuqin Lu and Tengfei Zhang and Ting Wang and Qian Wang and Qianqian Zhou},
  title   = {Artificial intelligence-oriented predictive model for the risk of postpartum depression: a systematic review},
  journal = {Frontiers in Public Health},
  year    = {2025}, volume = {13}, pages = {1631705},
  doi     = {10.3389/fpubh.2025.1631705}
}

@article{acog2023mentalhealth,
  author  = {{American College of Obstetricians and Gynecologists}},
  title   = {Screening and Diagnosis of Mental Health Conditions During Pregnancy and Postpartum: {ACOG} Clinical Practice Guideline No. 4},
  journal = {Obstetrics \& Gynecology},
  year    = {2023},
  volume  = {141},
  number  = {6},
  pages   = {1232--1261},
  doi     = {10.1097/AOG.0000000000005200}
}

@article{sharma2014peripartum,
  author  = {Sharma, Verinder and Mazmanian, Dwight},
title = {The {DSM-5} Peripartum Specifier: Prospects and Pitfalls},
  journal = {Archives of Women's Mental Health},
  year    = {2014},
  volume  = {17},
  number  = {2},
  pages   = {171--173},
  doi     = {10.1007/s00737-013-0406-3}
}

@article{wang2021ppdprevalence,
  author  = {Ziyi Wang and Jiaye Liu and Huan Shuai and Zhongxiang Cai and Xia Fu and Yang Liu and Xiong Xiao and Wenhao Zhang and Elise Krabbendam and Shuo Liu and Zhongchun Liu and Zhihui Li and Bing Xiang Yang},
  title   = {Mapping Global Prevalence of Depression Among Postpartum Women},
  journal = {Translational Psychiatry},
  year    = {2021},
  volume  = {11},
  number  = {1},
  pages   = {543},
  doi     = {10.1038/s41398-021-01663-6}
}

@misc{axios2026clancy,
  author       = {{Axios Boston}},
  title        = {Lindsay Clancy trial ends in mistrial},
  year         = {2026},
  month        = sep,
howpublished = {Axios},
  url = {https://www.axios.com/local/boston/2026/09/04/lindsay-clancy-trial-mistrial},
  note         = {Accessed: 2026-11-21}
}

\end{document}